\documentclass[aps,prb,showpacs]{revtex4}
\usepackage{amssymb}
\usepackage{graphicx}

\begin{document}

\title{Subgap tunnelling through channels of polarons and bipolarons in chain conductors.}
\author{S. I. Matveenko$^{1}$ and S. Brazovskii$^{2,1}$}
\affiliation{$^1$Landau Institute for Theoretical Physics, Kosygina Str. 2,
119334, Moscow, Russia\\
$^2$CNRS, B\^{a}t.100, Universit\'{e} Paris-Sud, 91405 Orsay, France.}
\date{}

\begin{abstract}
We suggest a theory of internal coherent tunnelling in the
pseudogap region where the applied voltage is below the free
electron gap. We consider quasi 1D systems where the gap is
originated by a lattice dimerization (Peierls or SSH effect) like
in polyacethylene, as well as low symmetry 1D semiconductors.
Results may be applied to several types of conjugated polymers, to
semiconducting nanotubes and to quantum wires of semiconductors.
The approach may be generalized to tunnelling in strongly
correlated systems showing the pseudogap effect, like the family
of High Tc materials in the undoped limit. We demonstrate the
evolution of tunnelling current-voltage characteristics from
smearing the free electron gap down to threshold for tunnelling of
polarons and further down to the region of bi-electronic tunnelling
via bipolarons or kink pairs. The interchain tunnelling is
described in a parallel comparison with the on chain optical
absorption, also within the subgap region.
\end{abstract}

\maketitle

\section{Introduction.}

The interchain, interplane transport of electrons in low
dimensional (quasi 1D, 2D) materials attracts much attention
\cite{sigma-c} in view of striking differences between
longitudinal and transverse transport mechanisms revealing a
general problematics of strongly correlated electronic systems.
Beyond the low field (linear) conduction, the tunnelling
current-voltage J-U characteristics $J(U)$, $\sigma=dJ/dU$ are of
particular importance. The interest has been renewed thanks to
recently developed \cite{latyshev} design of intrinsic tunnelling
devices where electronic transitions between weakly coupled chains
or planes take place in the bulk of the unperturbed material.

The first feature one expects to see at any tunnelling experiment
in gapful conductors is the regime of free electrons when the
current onset corresponds to the voltage $U=E_{g}^{0}$ of the gap
in the spectrum of electrons. But contrarily to usual systems,
like semi- or even superconductors, there is also a possibility
for tunnelling within the subgap region $E_{g}<U<E_{g}^{0}$. It is
related to the pseudogap (PG) phenomenon known for strongly
correlated electrons in general, well pronounces in quasi 1D
systems and particularly in cases where the gap is opened by a
spontaneous symmetry breaking (see \cite{mb:02} and refs.
therein). The PG is originated by a difference, sometimes
qualitative, between three forms of electronic states: a) short
living excitations which are close to free electrons, b) dressed
stationary excitations of the correlated systems, and c) added
particles which modify the ground state itself \cite{PGdef}. For
our typical examples of electrons on a flexible lattice, the
modification results in self-trapped states (b) like single
particle $\nu =1$ polarons with energies $W_{1}<\Delta_{0}$ below
the single electron (a) activation energy $\Delta _{0}$; then the
new gap $E_{g}<E_{g}^{0}=2\Delta_{0}$ will be observed as a true
threshold with the PG in between. There may be also contributions
of two-particle $\nu =2$ states (c) - bipolarons, which energy
gain per electron is larger than for polarons $W_{2}<2W_{1}$.
While the cases (a,b) are common for low symmetry and discrete
symmetry cases, for (c) there is a further drastic effect of a
spontaneous symmetry breaking like the case of the polyacethylene
$(CH)_{x}$ or of some doubly commensurate CDWs. Now the bipolarons
are decoupled into particles with a nontrivial topology, solitons
or kinks, changing the sign of the order parameter of the
dimerization. The situation is further intricate in systems with a
continuous GS degeneracy like Incommensurate Charge Density Waves
(ICDW) or Wigner crystals. Here even the self-trapping of a single
electron is allowed to lead to topologically nontrivial states,
the amplitude solitons ASs. In the same class we find a more
common case of acoustic polarons in a 1D semiconductor \cite{mb-ecrys,bm:03}.

Properties of systems with different types of the GS degeneracy,
and required theoretical approaches, are quite different. Here we
shall concentrate on systems with a discrete, precisely double,
degeneracy which also include most basic elements of
non-degenerate systems. Theoretically, the tunnelling in CDWs was
studied in details for regimes of free electrons \cite{free} when
the current onset corresponds to the voltage $U=E_{g}^{0}$ of the
gap in the spectrum of electrons. We shall consider the tunnelling
in the PG regime. We shall follow the method \cite{mb:02}
developed for studies of single particle spectral density
$I(p,\omega)$ in applications to PES and ARPES intensities. We
refer to this publication for details in techniques and
literature.

A word of notations. In the following we shall invoke many
quantities with the dimension of energy (or frequency, since we
shall keep $\hbar=1$) which will be classified according to
different characters (with indices). $U>0$ and $\Omega>0$ will be
the external voltage difference for tunnelling and the external
frequency for PES or optics. $E>0$ will always stay for electronic
eigenvalue in a given potential, negative values will be addressed
explicitly as $-E$. $V_{\nu}(E)$ will be branches of a total
energy (of deformations together with electronic energies)
supporting eigenstates $\pm E$ which may be filled with occupation
numbers $\nu=0,1,2$. $W_{\nu}$ will be total energies of
stationary states, that is
$W_{\nu}=\min_E V_{\nu}(E)=V_{\nu}(E_{\nu})$. $\omega_{0}\ll E$
will be the frequency of a collective mode (phonons specifically to CDWs)
which interaction with electrons is responsible for their
self-trapping. The collective deformation $\Delta(x)$ will also be
measured as the potential energy experienced by electrons. We
shall keep the electron charge $e=1$ hence potentials will be
measured as energies and the interchain current $J$ will have the
dimension \textit{number-of-particles/unit-time/unit-length}. The
indices $j=a,b$ will number coupled chains; indices $i$ will
number moments $\tau_{i}$ of time for virtual processes.

\section{Spectroscopies of the pseudogap.}

A possibility of tunnelling or of other excitations within the gap in
spectra of free electrons $E<\Delta_{0}$ is related to a more
general phenomenon of the pseudogap PG. For electrons, the PG
signifies the remnants of the spectral density $I(\Omega ,P)$, or
the integrated one $I(\Omega)=\int I(\Omega ,P)dP/2\pi $, at
$W_{1}<\Omega <\Delta_{0}$ where $W_{1}$ is the absolute boundary
of the spectrum. $W_{1}$ is the energy of a fully dressed state of
one electron interacting with other degrees of freedom. (There may
be totally external modes like deformations or polarizations for
usual polarons, external modes essentially modified by the bath of
electrons like in CDWs, internal collective modes of electronic
system itself like in SDWs.) Most commonly, the self-trapped state
of one electron is known as the "polaron " while more complex
objects, solitons, can appear for systems with continuously
degenerate GSs (see \cite{braz:84} for a review).

The functions $I(\Omega)$ and $I(\Omega ,P)$ are measured directly
in PES and ARPES experiments (these abbreviations stay for the
integrated Photo-Emission Spectroscopy and for the Angle (that is
momentum) Resolved one). As such they have been studied
theoretically for the PG region by the present authors
\cite{mb:02,mb-ecrys,bm:03} and we refer to these publications for
a more comprehensive discussion and for the literature
review. The one electron spectra can be accessed also in
traditional external tunnelling experiments: junctions or STM. For
the last case, and practically for macroscopic point junctions,
only the integrated $I(\Omega)$ is measured. Elements of the full
dependence $I(\Omega ,P)$ become necessary to describe strongly
anisotropic materials (layered quasi-2D or chain quasi-1D ones)
where the coherent tunnelling is realized in internal junctions of
"mesa" type devices \cite{latyshev}. Here the tunnelling goes
between adjacent layers within the single crystal of the same
material, hence the momentum is preserved. In a simpler version,
the internal subgap tunnelling takes place from free electrons of
some metallic bands or pockets to polaronic states within gapful
spectra which probably takes place in $NbSe_{3}$ \cite{latyshev+}.
Otherwise it measures actually the joint spectral density for
creation a particle-hole excitation at adjacent chains (the
interchain exciton). In this respect it will be instructive to
compare the coherent tunnelling and the subgap optical absorption
OA (see a short excursion and references in \cite{mb:02}, III.E). A less
expected version of the internal tunnelling is a possibility for
bi-electronic transfers (tunnelling of bipolarons or of
kink-antikink pairs) which usually is attributed only to
superconductors. We shall see that these processes extend the PG
further down to even lower voltages.

In any case, the tunnelling current $J(U)$ is given by the
transition rate of electrons between two subsystems $a,b$ kept at
the potential difference $U$. For a weak coupling $t_{\bot}$, the
electron tunnelling from $a$ to $b$ is given by the convolution of
spectral densities
\[
J\sim t_{\bot}^{2}\int_{0}^{U}I_{a}(-\Omega ,P)I_{b}(U-\Omega,P)
\left\vert \Lambda(P)\right\vert^{2}d\Omega dP
\]
if the momentum is conserved, or of their integrals $J\sim
t_{\bot}^{2}\int_{0}^{U}I_{a}(-\Omega)I_{b}(U-\Omega)d\Omega$ for
the incoherent tunnelling. (Everywhere we assume $T=0$.) Recall
that for free electrons with a spectrum $E(P)$ we have $I(\Omega
,P)\sim \delta (|\Omega|-E(P))$ while $I(\Omega)$ becomes the DOS
$I(\Omega)\Rightarrow N(\Omega)$, e.g. for $D=1$ $I(\Omega)\sim
t_{\bot}^{2}m^{1/2}(\Omega -\Delta_{0})^{-1/2}$ near the bottom of
the free band where the electron effective mass is $m$.

Consider briefly the case where one of reservoirs, say $\#b$, is
composed by free electrons with a known DOS $N_{b}(\Omega)$. One
of applications of a tunnelling between the free spectrum and the
PG may be the case of several families of conjugated polymers
(polypyrolle, polythiophene) where origins of filled, $\pi $ or
empty, $\pi^{\ast}$ bands below and above the gap are essentially
different. Then the polaronic effect, hence the PG, may be
pronounced only for one type of particles: electrons or holes. The
same concerns 1D systems made of semiconducting wires where both
effective masses and deformation potentials for electrons and
holes are usually very different. Then, for the incoherent
tunnelling, $I_{a}(U)$ gives directly either the tunnelling current
$J(U)\sim I_{a}(U)$ (if $N_{b}$ has a sharp peak at the Fermi
level, which is typical for using junctions with superconductors)
or the tunnelling differential conductance $\sigma =dJ/dU\sim
I_{a}(U)$ (if $N_{b}\approx const$ at the Fermi surface).

Below we shall be mostly interested in systems with the charge conjugated symmetry (or
qualitatively equivalent ones); the examples are carbon nanotubes,
 symmetric conjugated polymers like polyphenylenes, polyanilines
 and polymers where the gap is formed (partly at least) by the
 spontaneous symmetry breaking: the polyacethylenes (\cite{polymers}).
 Numerical details will be presented for the last rich case.
 In all these cases the PG\ will exist near both rimes $\pm\Delta_{0}$
  of the free excitation gap $E_{g}^{0}=2\Delta_{0}$.

Recall now some known results for $I(\Omega)$ within the PG
\cite{mb:02}. It has the form $I=A\exp (-S)$ where the action
$S=S(\Omega)$ is proportional to the big parameter of our
adiabatic approximation: $S\sim \Delta_{0}/\omega_{0}\gg 1$.
$S(\Omega)$ is determined by an optimal fluctuation localized in
space and time (an instanton) which supports the necessary
split-off local level $E$. In principle, the prefactor
$A=A(\Omega)$ also depends on $\Omega$ and may show power law
dependencies near extremals $0,W_{1}$. But within constraints of
the adiabatic approximation $\delta \Omega \gg \omega_{0}$ the
dependence $A(\Omega)$ is negligible in comparison with the one of
$S(\Omega)$. The characteristic value of $A$ may be important for
estimates of the overall magnitude of observable effects. Thus for
the single particle integrated intensity $A\sim (\omega
W_{1})^{-1/2}$ and
$A_{P}\sim(m\Delta_{0}\omega W_{1})^{-1/2}$ for the momentum resolved
intensity; here $m$ is the effective electron mass
$m\sim\Delta_{0}/v_{F}^{2}$. Appendix \ref{app:1} contains derivation
of the prefactor specifically for the tunnelling processes.

In limiting cases we have \cite{mb:02}

1. Near the entry to the PG, just below the free edge $\Delta_{0}$:
\begin{equation}
\Delta_{0}-W_{1}\gg \Delta_{0}-\Omega >0~:\
I=A\exp \left[ -\frac{cnst}{\omega_{0}}
\left(1-\frac{\Omega}{\Delta_{0}}\right)^{3/2}\right].
\label{Nh}
\end{equation}

\begin{figure}[tbp]
\begin{center}
\includegraphics[
trim=0.000000cm 0.000000cm -0.058507cm -0.040222cm, height=6cm, width=10cm
]{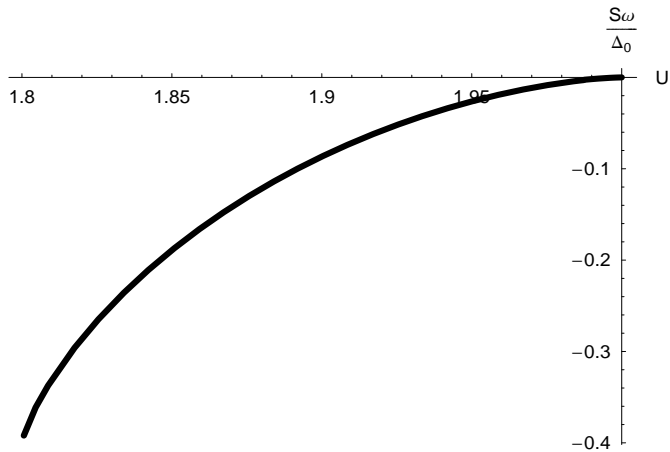}
\end{center}
\caption{The logarithmic plot for tunnelling or absorption
intensities $\ln I\sim -S$ in the pseudogap regime: between
$2W_{1}$ and $2\Delta_{0}$.} \label{fig:S4pol}
\end{figure}

2. Near the low end of the PG, just above the true spectral boundary $W_{1}$:
\begin{equation}
\Delta_{0}-W_{1}\gg \Omega -W_{1}>0~:\
I=A\exp \left[ -cnst\frac{\Delta_{0}-W_{1}}{\omega_{0}}-
cnst\frac{\Omega -W_{1}}{\omega_{0}}\ln \left(\frac{\Omega -W_{1}}{\Delta_{0}-W_{1}}\right) \right].
\label{Nl}
\end{equation}%
The total dependence $I(\Omega)$ and the values of numerical constants in the above
limiting laws, can be determined approximately \cite{mb:02} with the help of the instanton
techniques simplified by the zero dimensional reduction (the Anzats of an effective particle
which we shall recall and
extend below). The resulting curve is plotted at the figure \ref{fig:S4pol}.
Moreover, the regime (1.) can be mapped exactly \cite{mb:02} upon the problem of
a particle in a quenched random uncorrelated potential which here is created by
instantaneous quantum fluctuations of the media.
The known exact solution \cite{halperin} provides the reference value of the
coefficient in the exponent of (\ref{Nh}), from which our approximate value
differs only by $8\%$ \cite{mb:02}.

\begin{figure}[tbph]
\begin{center}
\includegraphics[
height=6cm, width=10cm
]{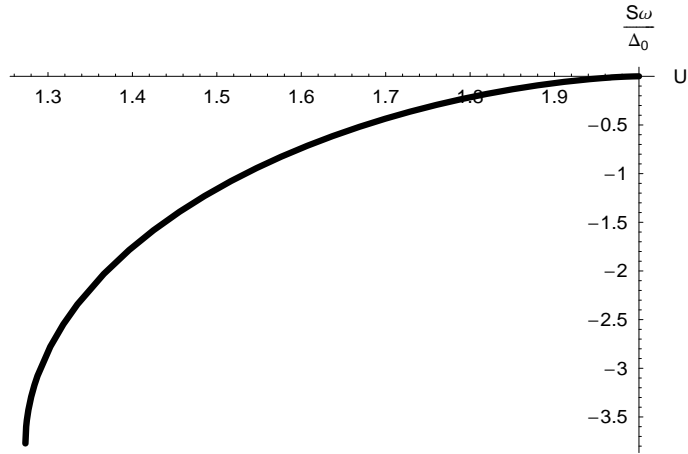}
\end{center}
\caption{The logarithmic plot for bi-electron tunnelling or for
optical absorption intensities $\ln I\sim -S$ in the two particle
pseudogap regime: between $2W_{s}$ and $2\Delta_{0}$. Notice the
$\protect\sqrt{\Omega -2E_{s}}$ singularity near the two soliton
threshold.} \label{fig:S4sol}
\end{figure}

Recall for comparison the usual regime $U>E_{g}^{0}$ of the
allowed tunnelling which is dominated by free electronic states.
The current of the coherent tunnelling between chains $a,b$ is given as
\begin{equation}
J\sim t_{\perp}^{2}\int \frac{dp}{2\pi}\delta(E_{b}(p)-U-E_{a}(p)) \left\vert
\Lambda(p)\right\vert^{2},\quad \Lambda(p)=\int dx\Psi_{bp}(x)\Psi_{ap}^{\ast}(x)
\label{free}
\end{equation}%
with $E_{a,b}(p)\approx \pm \left(\Delta_{0}+p^{2}/2m_{e}\right) $
and $\Psi_{jp}(x)$ being the Bloch functions. It is instructive to
compare the interchain tunnelling probability with the on chain
interband optical absorption OA when the matrix element of density
$\Lambda (p)$ changes to the one of the momentum: $t_{\bot}\Lambda
(p)\Rightarrow \Lambda_{OA}(p)$. In both cases the $e-h$ pair is
created and the same spectral densities are involved. The
difference is in matrix elements: the OA takes place between
states of opposite parity while the tunnelling requires for the
same parity. The on-chain OA between edges $\pm \Delta_{0}$ of the
free gap is known to be allowed since the parity of states near
$\pm \Delta_{0}$ is opposite, hence $\Lambda_{OA}$ is finite and
the OA intensity as a function of frequency $\Omega $ rises as
$I_{OA}\propto 1/\sqrt{\Omega -2\Delta_{0}}$. But for the same
reason, the tunnelling matrix element between identical chains is
prohibited at $p=0$ and the tunnelling will show only a weak edge
onset $J\sim \sqrt{\Omega -2\Delta_{0}}$. Nevertheless, in many
cases of gaps opened due to spontaneous dimerization, the
neighboring chains tend to order in antiphase. Now the shift by
half a period permute states with $E\gtrless 0$ then the parity of
states near opposite rims $\pm \Delta_{0}$ at neighboring chains
is equal, the tunnelling becomes allowed and the usual singularity
is restored: $J\propto 1/\sqrt{U-2\Delta_{0}}$.

Going down into the PG $\Omega<\Delta_0$, the above analysis applies to the on-chain
optics but changes drastically for the interchain tunnelling. The
tunnelling will be studied in details below, here we shall only
mention in advance an effect of spatial incoherence of optimal
quantum fluctuations at different chains which removes completely
the constraints of orthogonality. The case of the on-chain OA can
be analyzed briefly already here. The OA is given by the
convolution of two fast decaying functions of the energy
\begin{eqnarray}
I_{OA}\sim\int I(\Omega_{1},P)I(\Omega -\Omega_{1},P)
\left\vert\Lambda_{OA}(P)\right\vert^{2}d\Omega_{1}dP
\nonumber \\
\int\exp\left[ -S(\Omega_{1},P)-S(\Omega -\Omega_{1},P)\right]
\left\vert \Lambda (\Omega_{1},\Omega_{1}-\Omega)\right\vert^{2}d\Omega_{1}\sim
\nonumber\\
\Lambda^{2}(\Omega /2,-\Omega /2)\left(S^{\prime \prime}\right)^{-1/2}
\exp \left[ -2S(\Omega /2)\right]=
\nonumber\\
\left(S^{\prime \prime}(\Omega /2)\right)^{-1/2}
I^{2}(\Omega /2);\ S^{\prime \prime}=\frac{d^{2}S}{d\Omega^{2}}.
\label{IOA}
\end{eqnarray}%
Here we have used that for the convex function $S(\Omega)$, as given by (\ref{Nh},\ref{Nl}), the
minimum of the expression $S(\Omega_{1},P)+S(\Omega -\Omega_{1},P)$ lies at the middle
$\Omega_{1}=U/2$. At this point the electron levels $E$ and $E-\Omega $ are placed symmetrically,
wave functions have opposite parity, hence $\Lambda(E,-E)\neq 0$ is finite. This is the case of
typical Peierls insulators. But for systems where the basis wave functions of valent and conductive
bands have the same parity (the dipole OA is not allowed), $\Lambda _{OA}(E,-E)=0$ and we have to
consider in (\ref{IOA}) the deviations from the symmetry condition. Now
$\Lambda_{OA}(E_{a},E_{b})\sim (E_{a}+E_{b})^{2}$ and the saddle point integration in (\ref{IOA})
gives another factor of $1/S^{\prime \prime}$ which is small as $\sim \omega_{0}/\Delta_{0}$. We
arrive at the answer similar to (\ref{IOA}) but with the small prefactor $\left(S^{\prime
\prime}\right)^{-3/2}$.

Until now we did not consider the dependencies on the momentum $P$.
In the full range of $\Omega$ and $P$, the spectral function $I(\Omega,P)$
has a rich structure which can be tested in the ARPES experiments. In
observable quantities, the momentum dependence appears twice: via the matrix
element $\Lambda (P)$ and via the action $S(\Omega)\Rightarrow S(\Omega,P)$.
 The analysis is simplified for the regime 2: the low polaron boundary $W_1$.
Here the action dependence on $\Omega $ and $P$ comes through the
single variable $\Omega \Rightarrow \Omega +P^{2}/2M_{1}$ where
$M_{1}\sim m\Delta_{0}^{2}/\omega_{0}^{2}$, is a heavy mass of the
polaron center motion. This kinetic energy contribution can be
neglected in compare to the matrix element dependence on $P$ which
confines $\Lambda^{2}\sim \left\vert \Psi_{P}\right\vert^{4}$
within the characteristic momenta distribution $|\Psi_{P}|^{2}$ of
the wave function $\Psi (x)$ of the self-trapped electronic state
localized over the scale $L\sim \xi_{0}$: beyond $P\sim
\xi_{0}^{-1}=\Delta_{0}/\hbar v_{F}$, the function $\Lambda (P)$
falls off exponentially. (At this scale, the recoil kinetic energy
$P^{2}/2M\sim \omega_{0}^{2}/\Delta_{0}$ is small in compare to
the energy width $\epsilon \sim \left(S^{\prime
\prime}\right)^{-1/2}\sim
\left(\omega_{0}(W\,_{1}-\Omega)\right)^{1/2}\gg \omega_{0}$. Then
the final integration over $P$ affects only $\Lambda (P)$ and
gives a constant factor $\sim 1/\xi_{0}$.) Altogether we find for
tunnelling just the law (\ref{Nl}) with $\Omega \rightarrow U/2$.

In the regime 1., near the free edge, the states are shallow $\epsilon
=1-E/\Delta_{0}<<1$ and extended $L/\xi_{0}\sim \epsilon^{-1/2}\gg 1$.
The effective mass $M$ for the center of motion becomes light, energy
dependent
\[
M\sim \frac{\epsilon^{2}}{\omega_{0}^{2}L}\sim \frac{\epsilon^{5/2}}{\omega_{0}^{2}},
\]
but the characteristic energy scale of the form factor
$\sim M^{-1}L^{-2}\sim \omega_{0}^{2}\epsilon^{-3/2}$
is still small in comparison with the characteristic energy width $\sim\omega_{0}^{2/3}$
 of (\ref{Nh}). So again we integrate separately the factor
 $\int\Lambda^{2}(P)dP\sim\int dP\Psi_{P}^{4}\sim L$ to obtain an
 additional prefactor $\epsilon^{-1/2}$ for the tunnelling
law (\ref{Nh}) with $\Omega \rightarrow U/2$.

Recall that for the ARPES with independent variations of $\Omega $ and $P$,
their interference may lead to rather unexpected and potentially observable
phenomena (\cite{mb:02}, section III.D). One of them is the "quasi
spectrum": the intensity maximum over the line $\Delta_{0}-\Omega \sim
P^{1/2}$ within the PG $\Omega <\Delta_{0}$ (\cite{mb:02}, section III.D,
case B1, Eq.49). Another effect is the emergence of instantons at high $P$
within the domain of free electron region $>\Omega>\Delta_{0}$ leading to
the enhanced intensity within the band $0<\Omega -\Delta_{0}<cnst/P^{3}$
(\cite{mb:02}, section III.D, case B3, Eq.51).

\section{Tunnelling: the derivations.}

We shall follow the adiabatic method of earlier publications \cite{mb:02,mb-ecrys,bm:03}
assuming a smallness of collective frequencies $\omega_{0}$ in compare with the electronic gap:
$\omega_{0}\ll\Delta_{0}$. Now, electrons are moving in a slowly varying potential $\Delta(x,t)$,
so that at any instance $t$ their energies $E_{j}(t)$ and wave functions $\Psi_{j}(x,t)$
are defined from a stationary Schrodinger equation $H\Psi (x,t)=E(t)\Psi(x,E(t))$ (Eq. (\ref{dirac})
below will give an example). The Hamiltonian $H=H(x,\Delta(x,t))$ depends on the
instantaneous configuration $\Delta(x,t)$ so that $E(t)$ and $\Psi(x,E(t))$
depend on time only parametrically. Exponentially small probabilities which
we are studying here are determined by steepest descent paths in the joint space
 $[\Delta(x),t]$ of configurations and the time, that is by a proximity of the
 saddle point of the action $S$. It is commonly believed, in analogy with the
 usual WKB, that the saddle point, the extremum of $S$ over $\Delta$ and $t$,
 lie at the imaginary axis of $t$ so that, as usual, we shall assume $t\Rightarrow it$
 and correspondingly $S\Rightarrow iS$ since now on.

Consider the system of two weakly coupled chains $j=a,b$ which are put at
the electric potential difference $U$. The system is described by the
total action
\begin{equation}
S_{ab}=S_{a}+S_{b}+t_{\perp}\int dxdt
(\hat{\Psi}_{a}^{\dag}(x,t)\hat{\Psi}_{b}(x,t)+
\hat{\Psi}_{b}^{\dag}(x,t)\hat{\Psi}_{a}(x,t)),
\label{S-total}
\end{equation}%
where $S_{j}=S[\Delta_{j}(x,t)]$ are single chain actions and the term $\sim t_{\bot}$
describes the interchain hybridization of electronic sates. $\hat{\Psi}_{j}(x)$ are
operators of electronic states.\newline
The average transverse current is given by the functional integral
\begin{eqnarray}
J=\frac{\int D[\Delta_{j}(x,t)]~
it_{\perp}
(\hat{\Psi}_{a}^{\dag}(x,t)\hat{\Psi}_{b}(x,t)-
\hat{\Psi}_{b}^{\dag}(x,t)\hat{\Psi}_{a}(x,t))\exp [-S_{ab}]}
{\int dx D[\Delta_{j}(x,t)]\exp [-S_{ab}]}.
\label{J}
\end{eqnarray}

\subsection{One electron tunnelling.}

We consider first the processes originated by the transfer of one electron between the chains.
 They appear already in the first order of expansion of the exponent in (\ref{S-total})
 in powers of $t_{\perp}$, which contribution to the current (\ref{J}) can be written as
\begin{eqnarray}
J=Z_0^{-2}t_{\perp}^{2}\int D\left[ \Delta_{j}(x,t)\right] \int d(x-y)\int d(\tau_{1}-\tau_{2})
\\ \nonumber
[\Psi_{a}^{\ast}(x,\tau_{1})\Psi_{b}(x,\tau_{1}) \Psi_{a}(y,\tau_{2})\Psi_{b}^{\ast}(y,\tau_{2})
\exp(-S(\tau_{1}-\tau_{2},\Delta_{j}(x,t))),
\label{j1}
\end{eqnarray}%
where the normalizing factor $Z_0^{-2}$ is the denominator in (\ref{J}) taken at $t_{\perp}=0$.
 Here the time dependent action $S(\tau_{1}-\tau_{2})$ describes (in imaginary time)
 the process of transferring one particle from the doubly occupied level $E_{a}<0$
 of the chains $a$ to the unoccupied level $E_{b}>0$ of the chain $b$ at the time $\tau_{1}$
 and the inverse process at the time $\tau_{2}$. We have
\begin{eqnarray}
S(\tau_{1}-\tau_{2},\Delta_{j}(x,t)) &=&
\left\{\int_{-\infty}^{\tau_{1}}+\int_{\tau_{2}}^{\infty}\right\}
dt[L_{a}(0)+L_{b}(0)]+\int_{\tau_{1}}^{\tau_{2}}dt[(L_{a}(-1)+L_{b}(1)-U]
\nonumber\\
&=&
\int_{-\infty}^{\infty}dt[L_{a}(0)+L_{b}(0)]+
\int_{\tau_{1}}^{\tau_{2}}dt[(E_{b}+E_{a}-U],
\label{S(1,2)}
\end{eqnarray}%
\newline
where $L_{j}(\nu)=L([\Delta_{j}],\nu)$ are Lagrangians of the $j$-th
chain with the number of electrons changed by $\nu $. They are given as a
sum of the kinetic term and the potential $V_{\nu}$:

\begin{equation}
L_{j}(\nu)=\int dx\frac{(\partial_{t}\Delta)^{2}}{g^{2}\omega_{0}^{2}}+V_{\nu
}[\Delta (x,t)]~;\ V_{\nu}=V_{0}+|\nu |E .
\label{L(nu)}
\end{equation}
Here the potential term $V_{\nu}$ contains the energy of deformations and the sum
over electron energies in filled states $\alpha $ which include both the
vacuum states and the split off ones:
\begin{equation}
V_{_{\nu}}[\Delta (x,t)]=
\int dx\frac{\Delta^{2}}{2g^{2}}+\sum_{E_{\alpha}<E_{F}}E_{\alpha}[\Delta (x,t)]-W_{GS}
\label{V(nu)}
\end{equation}
(here $g$ is the coupling constant). $V_{\nu}$ is counted with respect to
the GS energy $W_{GS}$ so that in the non perturbed $\Delta\equiv\Delta_0$ state
$V_{_{\nu}}=|\nu |\Delta_{0}$ (the particle, electron for $\nu >0$ or hole
for $\nu <0$, added instantaneously to the non deformed GS is placed at the
lowest allowed energy, the gap rim $\Delta_{0}$).

The exact extremal (saddle point) trajectory is defined by equations

\begin{equation}
\delta S/\delta\Delta_{j}(x,t)=0~,\ \partial S/\partial\tau_{1}=
\partial S/\partial\tau_{2}=0  \label{dD,d1,d2}.
\end{equation}
Actually the explicit calculation of the action requires for
approximations. We shall follow a way \cite{mb:02} of the zero
dimensional reduction which reduces the whole manyfold of
functions $\Delta_{j}(x,t)$ to a particular class
\begin{equation}
\Delta_{j}(x,t)\Rightarrow\Delta_E(x-X_{j}(t),E_{j}(t)), \;
S[\Delta_{j}(x,t)]\Rightarrow S[E_{j}(t),X_j(t)]
\label{reduction}
\end{equation}
 of a given function $\Delta_E$ of $x$ (relative to a time dependent
center of mass coordinate $X_{j}(t)$). $\Delta (x)$ is
parameterized by a conveniently chosen (see \cite{mb:02} for
examples) parameter for which a universal and economic choice is
the eigenvalue $E_{j}(t)$. The requirement for the manyfold
$\Delta (x,E)$ is that it supports a pair of eigenvalues $\pm E$
split off inside the gap $(-\Delta_{0},\Delta_{0})$ which span the
whole necessary interval. The last simplification is to assume, in
the spirit of all approaches of optimal fluctuations
\cite{disorder,optimal}, that the potential supports one and only
one pair of localized eigenstates $\Psi(x,\pm E)$. Explicit
formulas for the Peierls case are given in the Appendix
\ref{app:2}.

Recall that for the OA problem we deal with one chain
characterized by one pair of functions $E(t)$ and $X(t)$. But for the
interchain tunnelling, the functions $E_{j}(t)$ at chains $j=a,b$
are not obliged to be identical and also the wells may be centered
around different points $X_{j}(t)$. Within such a
parametrization the variational equation in (\ref{dD,d1,d2})
yields the equation of motion for $E(t)$

\begin{equation}
f(E_{j})\left(\frac{dE_j}{dt}\right)^{2}-V_{\nu}(E_{j})-|\nu |U+H_{\nu j}~=0
~,\ \
f(E)=\frac{1}{g^2\omega_0^2}\int dx
\left(\frac{\partial \Delta (x,E)}{\partial E}\right)^{2},
\end{equation}%
where $H_{\nu j}=cnst$ are the Hamiltonians which must be constants within each
interval of integration in (\ref{S(1,2)}). Apparently, at the outer intervals
$(t<\tau_{1})$,$(\tau_{2}<t)$
$H_{0j}=0$ to provide the return to the GS with $V_{0}=0$ at $t\rightarrow \pm\infty $.
 At the inner interval $(\tau_{1}<t<\tau_{2})$ $H_{1j}=E_{j}(\tau_{1})+U=E_{j}(\tau_{2})+U$
 to preserve the continuity of velocities $\dot E_{j}$ at $t=\tau_{1,2}$.
 Since the values $E_{j}(\tau_{1,2})$ are
determined uniquely by the equation of motion at the outer intervals,
then $E_{j}(\tau_{1,2})$
coincide for both $j=a,b$, hence $H_{a}=H_{b}$ and the functions $E_{j}(t)$
become identical at any time $E_{a}(t)\equiv E_{b}(t)\equiv E(t)$.
(Still, the shapes are allowed to be shifted by different centers $X_{j}(t)$:
$\Delta_{a}(x-X_{a},t)\equiv \Delta_{b}(x-X_{b},t)$).
Finally the extremal conditions (\ref{dD,d1,d2}) with respect to impact
times $\tau_i$ in (\ref{S(1,2)}) yield
\[
\left. E_{a}(t)+E_{b}(t)\right\vert_{t=\tau_{1,2}}=U~~\mathrm{hence}~~
E_{a}(\tau_{1,2})=E_{b}(\tau_{1,2})=U/2 .
\]

The action is finite $S<\infty$, hence the transition probability is not
zero, only for a closed trajectory, that is at presence of a turning point
(as examples, see figures
\ref{fig:U12},\ref{fig:U14},\ref{fig:U18},\ref{fig:U16+C} in the
Appendix \ref{app:2}). There must be a minimal value of $E=E_{m}$ where
$\dot{E}=0$ hence $V_{1}(E_{m})=U/2$ and $E_{m}<U/2$. The last
condition requires for $minV_{1}(E)=W_{1}\leq U/2$
that is for $U>2W_{1}$ which determines the threshold voltage at
twice the polaron energy.

We arrive at the effective one chain problem with the doubled
effective action. The extremal tunnelling action is
$S_{tun}=2S_{I}$ which is twice the exponent appearing in the
spectral density $I$ with limiting laws (\ref{Nh},\ref{Nl}). The
full expression is
\begin{equation}
S_{tun}(U)=8\int_{E_{m}}^{U/2}dE\sqrt{f(V_{1}-U/2)}+8\int_{U/2}^{\Delta
_{0}}dE\sqrt{fV_{0}}~;\ V_{1}(E_{m})=U  \label{S(U)}.
\end{equation}

We obtain a final expression for the current after integration
over $\Delta_{j}(x,t)$ around the extremal taking into account the
zero modes related with translations of the instanton centers
positions $X_{j}(t)$. (Details of calculations are given in the
Appendix \ref{app:1})
\begin{equation}
J(U)\propto t_{\perp}^2 M_U \omega_0 \sqrt{\frac{dT}{dU}}
 \int\frac{dp}{2\pi}e^{-p^{2}l^{2}/4}
 \left\vert \Psi _{p}(U/2)\Psi_{p}(-U/2)\right\vert^{2}
 \exp[-2S_{I}(U/2)] ,
 \label{pre0}
\end{equation}%
where $\Psi_{p}$ is the Fourier transforms of the wave functions $\Psi (x)$,
the time $T$ is defined as  $ T = \int_{U/2}^{E_m} dE/ \dot{E}$.
 The mean fluctuational
displacement $l$ of the center of mass between the impact moments is given as
\[
l^{2}=\int_{\tau_{1}}^{\tau_{2}}\frac{dt}{M(E(t))}=2\int_{E_{m}}^{U/2}
\frac{dE}{M(E)}\frac{\sqrt{f(E)}}{\sqrt{V_{1}(E)-U/2}} ,
\]
where $M(E)$ is the translational mass:
\begin{equation}
M(E)=\frac{2}{g^{2}\omega_{0}^{2}}\int dx
\left(\frac{\partial \Delta (x,E)}{\partial x}\right)^{2}~,\ M_U=M(U/2).
\label{M(E)}
\end{equation}
Note that the prefactor in Eq. (\ref{pre0}), which is the matrix element between
orthogonal states $\Psi (E)$ and $\Psi (-E)$, is always nonzero due to the
integration over zero modes $X_{j}(t)$ (in contrast to results for the rigid
lattice where it obeys the selection rules); see more in the Appendix \ref{app:1}.

Comparing with the PES intensity $I(\Omega)$ calculated in
\cite{mb:02} we see that, up to pre-exponential factors, the
tunnelling current is proportional to the square of the PES
intensity $I$:
$
J\propto t_{\perp}^{2}I^{2}(\Omega =U/2)
$.
E.g. near the threshold $U=2W_{1}$ we can write
\begin{equation}
J\sim \frac{t_{\perp}^{2}}{\Delta_{0}\xi_{0}}
\left(\frac{\Delta_{0}}{\omega_{0}}\right)^{3/2}
\exp \left[ -C_{1}\frac{\Delta_{0}}{g\omega_{0}}\right]
\exp \left[ C_{2}\frac{(U-2W_{1})}{g\omega_{0}}\log \frac{2C_{3}\Delta_{0}}{(U-2W_{1})}\right].
\label{J1low}
\end{equation}
The coefficients $C_{i}\sim 1$ can be found numerically from (\ref{S(U)}) as
(for the Peierls model) $C_{1}=0.4$, $C_{2}=2.9$, $C_{3}=0.1$. (These values
differ from the corresponding ones in \cite{mb:02} because of different
normalizations of frequency $\omega_{0}$ in compare to $\omega_{ph}$).

\subsection{Bi-electronic tunnelling.}

It is known that the joint self-trapping of two electrons allows
to further gain the energy resulting in stable states different
from independent polarons. In general nondegenerate systems this
is the bipolaron, confined within the length scale twice smaller
than that of the polaron, the energy gain of the bound state
$\delta E=\Delta_{0}-E$ is four times that of the polaron and the
total energy gain of the bipolaron $\delta W_{2}=2\Delta
_{0}-W_{2}$ is also four times that of two polarons. (Certainly
these results neglect the energy loss due to the Coulomb repulsion
which may become critical for the stability of a shallow
bipolaron.) The same time, the total energy of one bipolaron
$W_{2}=2\Delta_{0}-\delta W_{2}$ is larger than the energy of one
polaron $W_{1}=\Delta_{0}-\delta W_{1}$ and even than the free
electron energy $\Delta_{0}$. This is why bipolarons cannot be
seen as thermal excitations while they are favored in case of
doping. The information on their existence comes from the ground
state of doped systems where bipolarons are recognized by their
spinless character and special optical features
(see \cite{bipol-polym} for experimental examples on conducting
polymers and \cite{bipol-th} for relevant theoretical models). An
important advantage of tunnelling experiments is a possibility to
see bipolarons directly, at voltages $U$ below the two-polaron
threshold $2W_{1}$ that is within the true single particle gap.
This possibility comes from the fact that, for bipolarons as
particles with the double charge $2e$, the voltage gain by
transferring from one chain to another is $2U$, hence the
threshold will be at $U=W_{2}<2W_{1}$. The probability of the
bi-electron tunnelling is small as it appears only in the higher
order $\sim t_{\perp}^{4}$ in interchain coupling. But it can be
seen as extending below the one-electron threshold where no other
excitations can contribute to the tunnelling current.

The bi-electronic contribution to the current can be written, by expanding
(\ref{S-total}) and (\ref{J}), as
\begin{eqnarray}
J_{2} &=& Z_0^{-2}t_{\perp}^{4}\int D[\Delta_a]D[\Delta_b]\int
\prod_{i=1}^{3}dy_{i}d\tau_{i}\exp \left(-S(\tau_{i})\right)
\nonumber\\
&&\left[ \Psi_{a}^{\ast}(x,\tau)\Psi_{b}(x,\tau)\Psi_{a}^{\ast}(y_{1},\tau_{1})
\Psi_{b}(y_{1},\tau_{1})\Psi_{b}^{\ast}(y_{b},\tau_{2})
\Psi_{a}(y_{2},\tau_{2})\Psi_{b}^{\ast}(y_{3},\tau_{3})\Psi_{a}(y_{3},\tau_{3}) -\{2\leftrightarrow
3\}\right], \nonumber
\end{eqnarray}%
which generalizes expressions (7) and (8) for the one
electron tunnelling. Here
\begin{eqnarray}
S(\{\tau_{i}\}) &=& \left\{\int_{-\infty}^{\tau}+\int_{\tau_{3}}^{\infty}\right\} dt[L_{a}(0)+L_{b}(0)]+
\nonumber\\
&&\left\{\int_{\tau}^{\tau_{1}}+\int_{\tau_{2}}^{\tau_{3}}\right\}
dt[(L_{a}(1)+L_{b}(1)-U]+\int_{\tau_{1}}^{\tau_{2}}dt[L_{a}(2)+L_{b}(2)-2U] .
\label{2e}
\end{eqnarray}
Within our model (\ref{L(nu)},\ref{V(nu)}) the potentials $V$ are additive in energy $E$, then the
action can be simplified as

\begin{eqnarray}
S(\{\tau_{i}\})=2\int_{-\infty}^{\infty}dtL_{a}(0,t)+   \nonumber\\
\left\{\int_{\tau}^{\tau_{1}}+\int_{\tau_{2}}^{\tau_{3}}\right\}
dt(E_{a}(t)+E_{b}(t)-U)+2\int_{\tau_{1}}^{\tau_{2}}dt(E_{a}(t)+E_{b}(t)-U) .
\end{eqnarray}

The extremal solution is defined, as above, by equations of the
type (\ref{dD,d1,d2}) but with four impact times $\tau_{i}$
instead of two. (Actually, in view of the time reversion symmetry,
the number of boundary conditions is twice smaller.) A similar
analysis of the extremal solution shows that optimal fluctuations
$\Delta_{j}(x,t)$ are identical in shape, up to shifts of their
centra: $X_{j}(t)$, $\Delta_{j}(x,t)\equiv \Delta (x-X_{j},t)$.
Hence the energies are identical $E_{a}(t)\equiv E_{b}(t)$, and
also the resonance conditions $2E(\tau_{i})=U$ take place at the
impact moments $\tau_{i}$. Moreover, the simple hierarchy of our
model $V_{2}-V_{1}\equiv V_{1}-V_{0}\equiv E-U/2$ shows that all
branches $V_{\nu}(E)$ cross at the same point $E=U/2$ (see figures
\ref{fig:U12},\ref{fig:U14},\ref{fig:U18} below). Then the
evolution $E(t)$ switches directly from the branch $\nu =0$ to the
branch $\nu =2$ and back, without following the intermediate
branch $\nu =1$. It means that the intervals $(\tau,\tau_{1})$ and
$\,(\tau_{2},\tau_{3})$ of one-electron transfers $\nu =1$ are
confined to zero: $\tau =\tau_{1},\,\tau_{2}=\tau_{3}$. In other
words, only processes of simultaneous tunnelling of pairs of
particles are left. Notice that this picture changes in more
general models, particularly taking into account important Coulomb
interactions. They add, to the energy branch of a shallow
bipolaron, the energy $\delta V_{2}\sim
(e^{2}/\epsilon_{\bot}L)\ln (L/a_{\bot})$ where $\epsilon_{\bot}$
is the dielectric susceptibility of the media in the interchain
direction, $L=L(E)$ is the localization length of $\Psi (x,E)$,
such that $E\sim 1/(mL^{2})$. Now the intermediate intervals
$(\tau =\tau_{1})$,$(\,\tau_{2}=\tau_{3})$ appear where the
evolution follows the $\nu =1$ branches, see figure
\ref{fig:U16+C}. With increasing Coulomb interactions this single
particle interval becomes more pronounced and the bipolaronic
threshold is shifted towards the one of two independent polarons.

In any case, the extremum solution for the action (\ref{2e}) is achieved on
the instanton trajectory given be the equation
\[f(E)\dot{E}^{2}=V_{U}(E)=\min\{V_{0}(E),V_{1}(E)-U/2,V_{2}(E)-U\}.
\]
The extremal action is
\begin{equation}
S_{2}(U)=8\int_{E_{m}}^{\Delta_{0}}dE\sqrt{f(E)V_{U}(E)}~; \
V_{U}(E_{m})=V_{2}(E_{m})-U=0  \label{S2(U)}.
\end{equation}
This action is finite if the turning point $E_{m}$ does exist, that is if
$U\geq\min V_{2}=W_{2}$.

Notice that, neglecting Coulomb interactions, the energy $V_{\nu}$
is determined only by the total number $\nu =\nu_{e}+\nu_{h}=\nu
(E)+(2-\nu (-E))$ of electrons and holes. Then the energy of the
bipolaron (both $\nu (E)$ and $\nu (-E)$ are either empty or
doubly occupied) and the energy of the exciton (both $\nu (E)=1$
and $\nu (-E)=1$ are singly occupied) are the same. Then the
trajectory of the bi-electronic tunnelling becomes the same as the
one for the case of optical absorption \cite{mb:02}, only the
action is doubled $S_{2}(U)=2S_{OA}(\Omega =U)$. Up to the
pre-exponential factor we have
\begin{equation}
J_{2}\propto t_{\perp}^{4}[I_{OA}(\Omega =U)]^{2},
\end{equation}%
where $I_{OA}(\Omega)$ is the optical absorption probability for one chain.

For common systems with a nondegenerate ground state, the dependence $S_{2}(U)$ resembles
qualitatively the law \ref{J1low} for the one-electron contribution,
with a similar behavior near the
threshold $U-2W_1\rightarrow U-W_2$. The situation changes for a
doubly degenerate ground state where the bipolaron dissolves into
a diverging pair of solitons (dimerization kinks). Thus for the
Peierls model the evaluation of (\ref{S2(U)}) gives, similar to the OA
law of \cite{mb:02}, near the two particle threshold
\begin{equation}
J_{2}\sim t_{\perp}^{4}\exp \left(-\max S_{2}+\frac{4}{g\omega_{0}}\sqrt{6\Delta_{0}(U-2W_{s})}\right)
~;\ \max S_{2}=C_{4}\Delta_{0}/g\omega_{0}
\label{J2min}
\end{equation}
with $C_{4}=3.77$. The overall dependence for the $\log J_{2}(U)\sim -S_{2}(U)$
is shown at the figure \ref{fig:S4sol}. Here we see explicitly that in the
order $\sim t_{\perp}^{4}$ the threshold voltage $U=2W_{s}$ is smaller than
$U=2W_{p}$ obtained in the order $\sim t_{\perp}^{2}$. Therefore this is the
main contribution to the current in the region $2W_{s}<U<2W_{p}$. Figure \ref{fig:S4sol}
 shows that the dependence of $J_2(U)$ near the low $U$ onset is much sharper than
 that of $J(U)$ at the figure \ref{fig:S4pol} near the polaronic onset which
 corresponds to the higher singularity in the limiting formula (\ref{J2min})
  in compare to (\ref{J1low}).

\section{Discussion and Conclusions.}

In quasi 1D systems with a gapful electronic spectrum, the
interchain tunnelling (as well as PES or OA) can be used to test
virtual electronic states within the pseudogap. Due to the
interaction of electrons with a low frequency mode, phonons in our
examples, the tunnelling is allowed in the subgap region
$U<E_{g}^{0}$ which forms the pseudogap. The one electron
processes lead to universal results similar both for systems with
the build-in gap and for those where the gap is due to the
spontaneous breaking of a discrete symmetry. The PG is entered
with the law (\ref{Nh}) and continues down to the threshold
$U_{1}=2W_{1}$, approached with the law (\ref{Nl}). This threshold
corresponds to the interchain transfer of fully dressed particles:
polarons with the energies $W_{1}$. But in tunnelling the PG is stretched even
further down thanks to processes of a simultaneous tunnelling of
two electrons. It terminates at the lower threshold $U_{2}=W_{2}$
or $U_{2}=2W_{s}$, $U_{2}<U_{1}$. Here $W_{2}$ is the energy of
the bipolaron - a bound state of two electrons selftrapped
together. In degenerate systems the bipolaron dissolves into
unbound solitons, hence the threshold at $2W_{s} $ with a more
pronounced dependence of the tunnelling rate (\ref{J2min}) as well
of the OA. Numerical results are presented at figures
\ref{fig:S4pol},\ref{fig:S4sol}.

There is an important difference between subgap processes and the
usual overgap transitions at $U,\Omega >E_{g}^{0}$ of free
electrons in a rigid system. It comes, beyond intensities, from
different character of matrix elements. Actually within the PG
region there are no particular selection rules since the wave
functions of virtual electronic states split off within the gap
are localized having a broad distribution of momenta. Then the PG
absorption is allowed independent on the interchain ordering.
Contrarily, the regular tunnelling across the free gap shows
an expected DOS singularity $\sim (E-E_{g}^{0})^{-1/2}$ for the
out of phase interchain order while for the in-phase order the
threshold is smooth $\sim (E-E_{g}^{0})^{1/2}$. This difference
may be important to choose an experimental system adequate for
studies of PGs. The smearing of the free edge singularity is a natural
criterium for existence of the PG below it \cite{Kim:93}. But the total
absence of this strong feature in systems with forbidden overgap
transitions can allow for a better resolution of the whole PG
region, down to the absolute threshold. Probably a very smooth
manifestation of gaps in usual tunnelling experiments on CDWs
\cite{tunneling}, while the gaps show up clearly through
activation laws, is related to this smooth crossover from the
overgap to the subgap region. (Notice that the existing experiments refer
mostly to ICDWs which, with their continuous degeneracy of the GS, must be
 studied specially which is beyond the scope of this article.)

Finally we shall discuss relations with other theoretical approaches. Most
theories of tunnelling, see \cite{free}, keep the following assumptions:
i. They refer to the overgap region where interactions or fluctuations are not
important and usually are not taken into account.
ii. They refer to the incoherent tunnelling, local in space, which is a usual
circumstance of traditional experiments. The PG in tunnelling was considered
by Monz et al in \cite{free} in the framework of the approach \cite{sadovskii}.
This method became popular recently in theories of the PG thanks to its easy
implementation: it is sufficient to average results for a rigid system over
a certain distribution of the gap values. Apparently this is the way to
describe an average over a set of measurements performed on similar systems
with various values of the gap, e.g. manipulating with the
temperature, the pressure or a composition. But actually, as we could see
above, the PG is formed by fluctuations localized both in space and time, the
instantons, with localization parameters depend on the energy deficit being
tested. There is an intermediate approach applied \cite{ohio} to a complex of
the PG phenomenon from optics to conductivity and susceptibility. It treats
fluctuations as an instantaneous disorder due to quantum zero point
fluctuations of the gap. Indeed, this picture can be well applied, as it was
done already in \cite{brazov:76}, but only to dynamical processes and only
in the upper PG region, just below the free gap $E_{g}^{0}$, which leads to
the law (\ref{Nh}). But deeper within the PG, the fluctuations are not
instantaneous: they require for an increasingly longer time and become
self-consistent with the measured electronic state leading to another
law and to appearance of the lower threshold. Generalizations and deeper
analysis of the model of the instantaneous disorder lead to interesting
theoretical studies \cite{monien}, but their applicability is very limited
unless the variable time scale is realized as we have demonstrated in this
and preceding articles.

Our approach can be compared to the work \cite{maki} on the fluctuational
creation of pairs of phase solitons in a 1D commensurate CDW under the
longitudinal electric field. But in our case me deal, in effect, with the
interchain tunnelling of pairs of solitons under the transverse field; also the
solitons have a more complex character of a multielectronic origin.

In conclusion, the presented and earlier \cite{mb:02,bm:03,mb-ecrys} studies
recall for the necessity of realizing the variable time scale of subgap processes
both in theory and in diverse interpretations of different groups of experiments
(dynamic, kinetic, thermodynamic) which address excitations with very different life times.

\acknowledgments S. M. acknowledges the hospitality of the Laboratoire de
Physique Th\'{e}orique et des Mod\`{e}le Statistiques, Orsay and the support
of the CNRS via the ENS - Landau foundation.

\appendix

\section{\label{app:2} Details on self-trapping branches.}

\begin{figure}[tbph]
\begin{center}
\includegraphics[height=5.5cm]{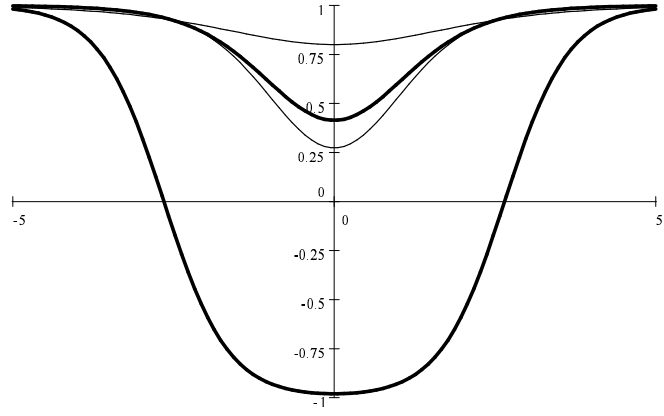}
\end{center}
\caption{Exact shapes $\Delta(x,E)$ of the equilibrium polaron, $E=2^{-1/2}$
(upper thick line) and of a nearly formed ($E=0.01$) pare of solitons (lower
thick line). Thin lines show exact shapes of optimal fluctuations necessary to
create these states by tunnelling.} \label{fig:shapes}
\end{figure}

We consider the system of weakly coupled dimerised chains. Each chain is
described by a usual electron-phonon Hamiltonian (Peierls, SSH models).
Electron levels $E$ and wave functions $\Psi =\Psi (x,E)$ are determined by
equations
\begin{equation}
[-v_F i\partial_{x}\sigma_{3}+\Delta(x)\sigma_{1}]\mathbf{\Psi}=E\mathbf{\Psi},
\label{dirac}
\end{equation}
where $\sigma_{1,3}$ are the Pauli matrices,
$\mathbf{\Psi}=(\Psi _{+},\Psi_{-})$, $\Psi_{\pm}(x)$ are the components
of electron wave functions near Fermi points $\pm p_{F}$, and the real
function $\Delta (x)$ is the amplitude of the alternating dimerization potential.
 The ground state of each chain is the Peierls dielectric with the gap $2\Delta_{0}$.
 The electron spectrum has the form
$E_{p}^{2}=v_{F}^{2}p^{2}+\Delta_{0}^{2}$ (in the following we shall put the Fermi velocity
$v_{F}=1$, $\Delta_{0}=1$ and, as everywhere, the Plank constant $\hbar =1$).
The excited states are
solitons (kinks), polarons and bi-solitons (kink-antikink pairs) which are
characterized by electron levels localized deeply within the gap
(see the review \cite{braz:84}). The one parametric family of configurations
$\Delta(x,E)$ supporting the single split-off pair of levels $\pm E$ can be written as
\begin{equation}
\Delta (x,E)=1-\frac{2\left(1-E^{2}\right)}{1+E\cosh \left(2x\left(1-E^{2}\right)^{1/2}\right)}
\label{bs}
\end{equation}
evolving from a shallow potential well at $E\approx 1$ through the stationary
configuration for a polaron $\nu =1$ to the pair of diverging kinks at
$E\rightarrow 0$ as shown at the figure \ref{fig:shapes}. The potentials
$V_{\nu}$ (for the $E$ level filling $\nu $) as functions of $E$ are given as
\begin{equation}
V_{\nu}(E)=\nu E+\frac{4}{\pi}\sqrt{1-E^{2}}-\frac{4}{\pi}E\cos^{-1}E .
\label{vbs}
\end{equation}
The translational mass can be found as
\begin{equation}
M(E)=\frac{8\Delta_{0}^{3}}{g^{2}\omega_{0}^{2}}
\left[\frac{1}{3}\tan^{3} \left(\cosh^{-1}\frac{1}{E}\right)
-E^{2}\cosh^{-1}\frac{1}{E}+E^{2}\sqrt{1-E^{2}}\right].
\end{equation}

Consider the matrix element between levels $\pm E$ in the Peierls
state. The wave function has two components $(u,w)$ according to
$\Psi =\Psi(x,E)=u\cos p_{F}x+w\sin p_{F}x$. Explicit expressions
for split-off states are $u,w\sim \sqrt{1-\Delta {}^{2}\pm
\partial_{x}\Delta}$. The equation for the bound eigenstate
(\ref{dirac}) shows the following symmetry: $w(x,E)=u(-x,E)$, $w(x,-E)=u(x,E)$,
$u(x,-E)=-u(-x,E)$. Then $\Psi (x,E)=((u(x),u(-x))$,
$\Psi(x,-E)=(-u(-x),u(x))$, with $u=u(x,E)$, which demonstrates explicitly the
orthogonality of $\Psi (x,E)$ and $\Psi (x,-E)$. The matrix element in Eq.
(\ref{pre0}) becomes
$\Lambda_{p}^{2}\sim\left\vert\Psi_{p}(E)\Psi_{p}(-E)\right\vert^{2}
=\left\vert-u_{p}u_{-p}+u_{p}u_{p}\right\vert^{2}$. At $p=0$,
$\Lambda=\Lambda_{0}=0$, hence for identical chains the transition at the free
gap $\Omega=2\Delta_{0}$ is forbidden which removes the singularity at the gap
threshold in a rigid system. But the true threshold at $2W_{1}$ for the subgap
absorption or tunnelling are not subjected to this selection rule since the wave
functions of localized states associated to the optimal fluctuation are
distributed over the momentum region $p\sim \xi_{0}^{-1}$.

Figure \ref{fig:shapes} shows exact shapes $\Delta (x,E)$ of the equilibrium
polaron (upper thick line) and of a well formed $(E=0.01)$ pair of solitons
(lower thick line). Thin lines show exact shapes of optimal fluctuations
necessary to create these states by tunnelling. Notice the much less pronounced
shapes for optimal fluctuations in compare to the final states which
facilitates the tunnelling.

\begin{figure}[ptbh]
\begin{center}
\includegraphics[height=5.5cm]{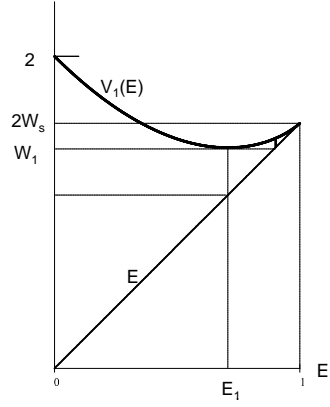}
\end{center}
\caption{Total energy of the single particle branch $V_{1}(E)$ as a function
of the  energy $E$ of the associated bound state.}
\label{fig:V1}
\end{figure}

Figure \ref{fig:V1} plots the total energy $V_{1}(E)$ of the single particle
branch as a function of the associated energy of the bound state. $E=1$,
$V_1(1)=1$ corresponds to the particle added to the unperturbed ground state,
at the bottom of the continuous spectrum. $E=0$ is the mid-gap state reached
for the limit of two divergent solitons when the total energy approaches the
maximal value $V_{1}(0)=2W_{s}=4/\pi\approx 1.27$. In between, at
$E_1=2^{-1/2}\approx 0.7$, $V_{1}(2^{-1/2})=W_{1}=2^{3/2}/\pi\approx0.9$, the
minimum corresponds to the stationary polaronic state. The short thin vertical
line between plots $E$ and $V_{1}(E)$ points to the configuration (upper thin
curve at the figure \ref{fig:shapes}) of the fluctuation necessary for
tunnelling to the polaron (the minimum of $V_{1}(E)$, upper thick curve at the
figure \ref{fig:shapes}).

\begin{figure}[ptbh]
\begin{center}
\includegraphics[height=5.5cm]{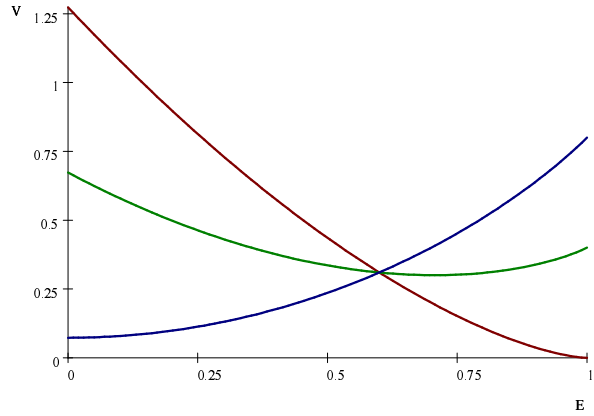}
\end{center}
\caption{Total energies $V_{\protect\nu}-\protect\nu U$ for different tunnelling
branches $V_{\nu}-\nu U$ as a function of the energy $E$ of the associated
bound state. This figure corresponds to $U=1.2$ which is below the
bi-electronic threshold.} \label{fig:U12}
\end{figure}

Next three figures plot the total energies $V_{\nu}-\nu U$ for different
branches as a function of the energy $E$ of the associated bound state (all in
units of $\Delta_{0}$). Branches are distinguished by their ordering at $E=1$.
Figure \ref{fig:U12} corresponds to the potential $U=1.2$ which is below
the bi-electronic threshold; no branch is crossing $V=0$ axis, hence no final
action is allowed and the current is zero.

\begin{figure}[ptbh]
\begin{center}
\includegraphics[height=5.5cm] {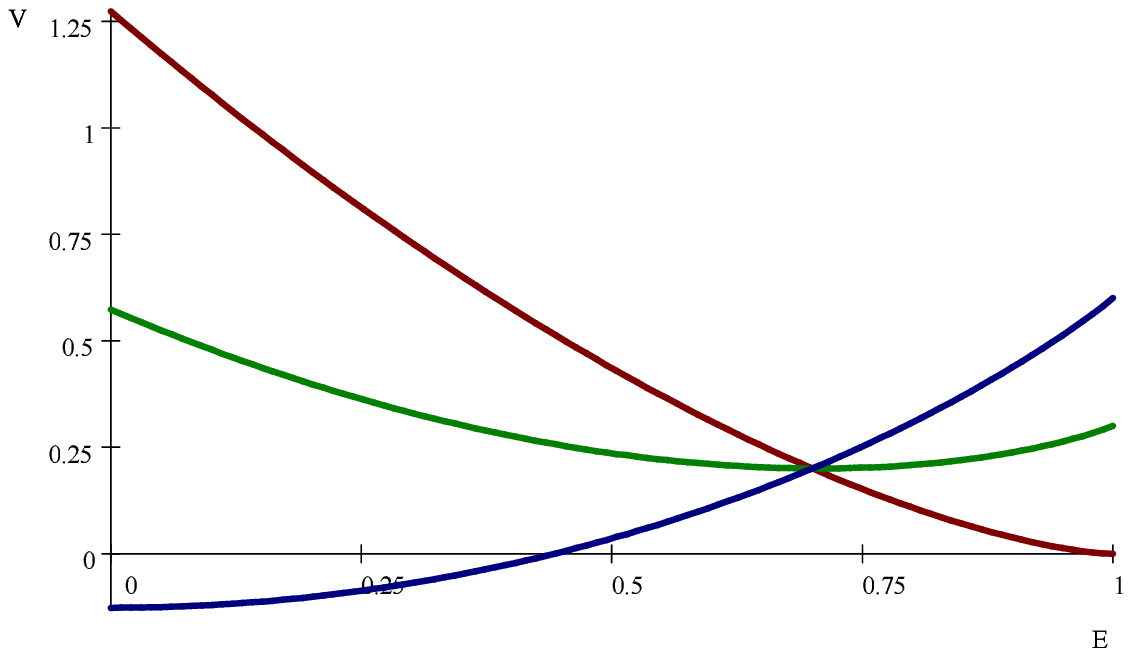}
\end{center}
\caption{Tunnelling branches $V_{\nu}-\nu U$ as a function of the energy $E$ of
the associated bound state. This figure corresponds to $U=1.4$ which is between
the thresholds for tunnelling of bipolarons and polarons.} \label{fig:U14}
\end{figure}

Figure \ref{fig:U14} corresponds to the potential $U=1.4$ which is between the
bi-electronic threshold $2W_{s}=4/\pi\approx1.3$ and the polaronic one
$2W_{1}=1.8$; the bi-electronic branch crosses the axis $V=0$ at the point
$E_{m}$, the action is finite, hence a nonzero tunnelling of two electrons is
allowed.

\begin{figure}[ptbh]
\begin{center}
\includegraphics[height=5.5cm]{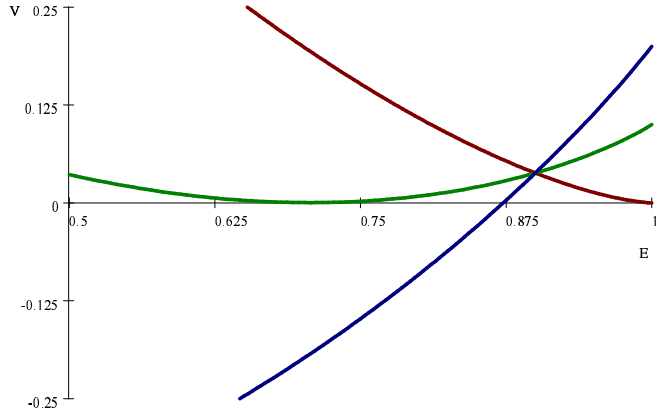}
\end{center}
\caption{Tunnelling branches $V_{\protect\nu}-\protect\nu U$ as a function of
the energy $E$ of the associated bound state (over a selected interval). This
figure corresponds to $U=1.8$ which is just at the polaronic threshold, above
the bipolaronic one.} \label{fig:U18}
\end{figure}

Figure \ref{fig:U18} corresponds to the potential $U=1.8$, above the
bi-electronic threshold $2W_{s}=4/\pi \approx 1.3$, exactly at the polaronic
one $2W_{1}=1.8$. Now two parallel processes of one- and two- electron
tunnelling are allowed.

\begin{figure}[tbh]
\begin{center}
\includegraphics[height=5.5cm]{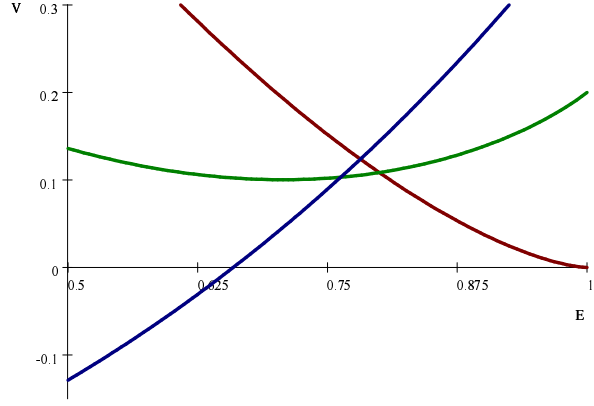}
\end{center}
\caption{Tunnelling branches $V_{\protect\nu}-\protect\nu U$ as a function of
the energy $E$ of the associated bound state.
This figure corresponds to $U=1.6$ which is between the thresholds for
tunnelling of bipolarons and polarons. Contrarily to the figure
\protect\ref{fig:U14}, the Coulomb interaction is taken into account which
lifts the crossing degeneracy.} \label{fig:U16+C}
\end{figure}

The figure \ref{fig:U16+C} corresponds to the potential $U=1.6$ between the
bi-electronic threshold $2W_{s}=4/\pi \approx 1.3$, and the polaronic one
$2W_{1}=1.8$. Contrary to the figure \ref{fig:U14}, the Coulomb interaction is
taken into account which lifts the degeneracy of the earlier crossing point of
three branches. The one electron term $\nu =1$ does not cross $V=0$ axis yet,
but it passes below two other terms in a vicinity of their crossing. Now the
optimal bi-electronic tunnelling takes place via a sequence of two single
electronic processes confined in time.

\section{\label{app:1} Derivation of prefactor}

We need to perform the integration over $\Delta_{j}(x,t)$ around
the extremal taking into account the zero modes related to
translations of positions $X_{j}(t)$ of the instanton centers. The
path integration over the gapless mode $X(t)$ is important,
particularly for the matrix element: the overlap of wave functions
evolves following $X_{a}(t)-X_{b}(t)$ while their localization
follows the evolution of $E(t)$. We shall work within the zero dimensional
reduction of Eq.(\ref{reduction}).

We expand the field $\Delta_{j}(x,t)$ in the vicinity of  the
instanton solution as
\begin{equation}
\Delta_{j}(x,t) = \Delta_0(x-X_j (t),E_j(t) + \delta( x-X_j(t),t)).
\end{equation}

Following \cite{mb:02}, we rewrite (\ref{j1}) as

\begin{eqnarray}
J\propto t_{\perp}^{2}\prod_{j=a,b}\int d(x-y) d(\tau_1-\tau_2)
\int D[E_j]D[X_j]J_{X_j}J_{E_j}\exp(-S)\nonumber\\
 \Psi_{a}^{\ast}(x-X_{a}(\tau_1),E(\tau_1))\Psi_{a}(y-X_{a}(\tau_2),E(\tau_2))
 \Psi_{b}(x-X_{b}(\tau_1),-E(\tau_1))\Psi_{b}^{\ast}(y-X_{b}(\tau_2),-E(\tau_2)),
\label{jjj}
\end{eqnarray}
where  $J_{X}=\propto\prod_{n=1}^{N}\sqrt{M(E(t_n))}$,
$J_{E_{j}}\propto\prod_{n=1}^{N}\sqrt{f(E_{j}(t_n))}$ are the Jacobians of the
transformation (\ref{reduction}). ($N\rightarrow\infty$ is the number of points
for the intermediate discretization of the time axis.) We integrate over the
zero mode $X(t)$ and take into account  fluctuations of the instanton shape due
to  variations of the parameter $E_0 (t)$.

The action in (\ref{reduction}) has the form
\[
S[X,E]=\sum_{j=a,b}dt\left(M(E_{j}(t))\dot{X_{j}}^{2}/2+f(E_{j})\dot{E}_{j}^{2}/2
+{V}_U(E_j)\right)
\]
with $V_U(E)$ from (\ref{S2(U)}). The integration over $DX_i(t)$ is
carried out exactly after the transformation $M \dot{X}^2=\dot{Z}^2$ using
the known expression

\begin{equation}
\int
D[x]\exp\left(-\int_{t_{1}}^{t_{2}}dt(\frac{\dot{x}^{2}}{2}+V(x))\right)
\sim\exp(-S_{cl})\sqrt{\frac{d^{2}S_{cl}}{dx_{1}dx_{2}}},
\label{dashen}
\end{equation}
where $x_{1}=x(t_{1})$, $x_{2}=x(t_{2})$.
Next, we perform in (\ref{jjj}) the remnant integrations over coordinates at the impact moments:
$X_1 =X_a(\tau_1)$, $X_2=X_a(\tau_2 )$, $Y_1 =X_b(\tau_1 )$, $Y_2 =X_b(\tau_2)$:

\begin{eqnarray}
J \propto t_{\perp}^{2}\int dxd\tau_{1}dX_{1}dX_{2}e^{-\frac{(X_{1}-X_{2})^{2}
}{l^{2}}}\Psi_{a}^{\ast}(x-X_{1},E(t_{1}))\Psi_{a}(y-X_{2},E(t_{2}))
\frac{\sqrt{M_{1}}}{l_{1}} \nonumber\\
dY_{1}dY_{2}e^{-\frac{(Y_{1}-Y_{2})^{2}}{l^{2}}}
\Psi_{b}^{\ast}(x-Y_{1},-E(t_{1}))\Psi_{b}(y-Y_{2},-E(t_{2}))\frac{\sqrt{M_{2}}}{l_{2}} \exp[-S_E(E)].
\end{eqnarray}
Here $M_i=M_i(\tau_{1,2})$, and the same for $l_i$, are functions of energies
in these points which finally become $E_i=E_i(\tau_{1,2})=U/2$.
 Using Fourier transforms, we rewrite the product of wave functions as
\[
\int dpdqd\tilde{p}d\tilde{q}\Psi_{a,p}^{\ast}e^{-ip(x-X_{1})}\Psi_{a,q}e^{iq(y-X_{2})}
\Psi_{b,\tilde{p}}e^{i\tilde{p}(x-Y_{1})}\Psi_{b,\tilde{q}}e^{-i\tilde{q}(y-Y_{2})}.
\]
Integration over $dxdy$ gives $\delta(p-\tilde{p})$, $\delta(q-\tilde{q})$ and
integration over $dX_{2},dY_{2}$ gives $L\delta(p-q)$. After
integration over $d(X_{1}-X_{2})d(Y_{1}-Y_{2})$ we arrive at the result (\ref{pre0}).
The factor $\sqrt{{dT}/{dU}}=1/\sqrt{d^{2}S/dT^{2}}$ in (\ref{pre0})  after
integration over $D[E_i(t)]$ which was performed using again the equation (\ref{dashen}).


\begin{thebibliography}{99}
\bibitem{PGdef} We view the PG as a partial filling of an expected clear spectral gap,
which is adequate to CDW
physics. Another view, typical in the field of Hihg-Tc
superconductors, is that the PG is a suppressiion of the expected
metalic DOS. These too approaches might be convergent.
\end{thebibliography}
\end{document}